\documentclass[11pt]{article}
\usepackage{amsmath}
\usepackage{amssymb}
\usepackage{graphicx}
\usepackage[mathscr]{eucal}
\usepackage[usenames]{color}
\usepackage[latin5]{inputenc}
\usepackage{url}
\usepackage{authblk}

\numberwithin{equation}{section}

\newcommand{\be}{\begin{equation}}
\newcommand{\ee}{\end{equation}}

\begin{document}
\clearpage
\thispagestyle{empty}

\title{\bf Heisenberg-type higher order symmetries of superintegrable systems separable in cartesian coordinates}

\author[1]{F. G\"ung\"or\thanks{gungorf@itu.edu.tr}}
\author[2]{\c{S}. Kuru\thanks{sengul.kuru@science.ankara.edu.tr}}
\author[3]{J. Negro\thanks{jnegro@fta.uva.es}}
\author[3]{L.M. Nieto\thanks{luismiguel.nieto.calzada@uva.es}}

\affil[1]{Department of Mathematics, Faculty of Science and Letters,
Istanbul Technical University, 34469 Istanbul, Turkey}
\affil[2]{Department of Physics, Faculty of Science, Ankara
University, 06100 Ankara, Turkey} 
\affil[3]{Departamento de
F\'{\i}sica Te\'{o}rica, At\'{o}mica y \'{O}ptica, Universidad de
Valladolid, 47011 Valladolid, Spain}

\date{\today}

\maketitle

\begin{abstract}
Heisenberg-type higher order symmetries are studied for both
classical and quantum mechanical systems separable in cartesian
coordinates. A few particular cases of this type of superintegrable
systems were already considered in the literature, but here they are
characterized in full generality together with their integrability
properties. Some of these systems are defined only in a region of
$\mathbb R^n$, and in general they do not include bounded solutions.
The quantum symmetries and potentials are shown to reduce to their
superintegrable classical analogs in the $\hbar \to0$ limit.
\end{abstract}


\newpage



\section{Introduction}

We will consider a classical two--dimensional Hamiltonian, $H$,
separable in
Cartesian coordinates having the form
\begin{equation}\label{hc}
H=H_x+H_y,\qquad H_x=p_x^2+v_x(x),\qquad H_y=p_y^2+v_y(y)\,.
\end{equation}
As this system is separated, there are two integrals of
motion: one of them is the Hamiltonian itself $H$, while the other one, $A$, can be
taken, for example, as the difference of both component Hamiltonians,
$A= H_x - H_y$. Therefore, the system is integrable (there are two constants of motion
$H$ and $A$ in involution). In this paper, we want to search for systems having
this general structure and allowing for another
independent constant of motion, $B$, polynomial in the momentum variables $p_x,p_y$.
So, such systems will be superintegrable, with three independent integrals $H,A$ and $B$.
We will restrict ourselves to a special class of such
superintegrable systems, based on a particular property of the one--dimensional
component Hamiltonians, as it is shown below.

First, we want that the additional integral of motion
$B_n$ for this Hamiltonian be also separable in
the coordinates $x,y$ in the form
\begin{equation}\label{bc}
B_n(x,y,p_x,p_y)=B_{nx}(x,p_x)-B_{ny}(y,p_y).
\end{equation}

Second, we ask the functions $B_{nx}$ and $B_{ny}$ to be $n$--degree
polynomials in the momentum variables $p_x, p_y$:
\begin{equation}\label{bcx}
B_{nx}(x,p_x)=\sum_{j=0}^n {f_j(x)\,p_x^{j}},\qquad B_{ny}(y,p_y)=\sum_{j=0}^n {g_j(y)\,p_y^{j}}\,,
\end{equation}
the coefficients $f_j(x)$, $g_j(y)$ being some unknown functions, depending on the variables $x$ and $y$, respectively.

Third,  the functions $B_{nx}$ and $B_{ny}$ must satisfy the following
Heisenberg-type Poisson brackets (PB):
\begin{equation}\label{bhcxy}
\{H_x,B_{nx}(x,p_x)\}=\{H_y,B_{ny}(y,p_y)\}={\rm constant}=1.
\end{equation}
The constant can be taken, without loss of generality, equal to one.
Then, it is clear that the function $B_n$ given by (\ref{bc}) will
satisfy, together with the Hamiltonian (\ref{hc}), the following PB,
\begin{equation}\label{bhc}
\{H,B_n(x,y,p_x,p_y)\}=\{H_x,B_{nx}(x,p_x)\}-\{H_y,B_{ny}(y,p_y)\}=0\,.
\end{equation}
In this way, we have arrived to an `extra' constant of motion to
achieve superintegrability. Such a constant of motion is called to
be of Heisenberg type, since it is based on the Heisenberg algebra
(\ref{bhcxy}) for each of the one dimensional components: $\langle
H_x,B_{nx},1 \rangle$ and $\langle H_y,B_{ny},1 \rangle$. Each one
dimensional Hamiltonian $H_x$, $H_y$ is called Heisenberg
Hamiltonian. Depending on the value of $n$ we will speak of
$n$-degree superintegrable system, when $n\geq 3$  the constant of
motion $B_3$  is said to be of `higher order' (since the `standard'
constants of motions are of degree two). Recall that the PB  for the
functions $F(x,y,p_x,p_y),\,G(x,y,p_x,p_y)$ is defined in the usual
way,
\begin{equation}\label{pb}
\{F ,G \}=\frac{\partial F}{\partial x}\frac{\partial G}{\partial p_x}+\frac{\partial F}{\partial y}\frac{\partial G}{\partial p_y}-\frac{\partial F}{\partial p_x}\frac{\partial G}{\partial x}-\frac{\partial F}{\partial p_y}\frac{\partial G}{\partial y}\,.
\end{equation}
We will see that some of the superintegrable Hamiltonians are only defined in regions
of the plane ${\mathbb R}^2$, furthermore the corresponding potentials
will not allow for a bounded motion. Therefore, the evolution of a particle in these
potentials will have only a piece of its trajectory in the domain of
superintegrability. We could `extend' any such Hamiltonian to another one defined
in the whole plane, but it will not be superintegrable anymore, and in case
this new extended Hamiltonian has a bounded motion, in general this motion will not be
periodic.

This program can also be carried out for the corresponding quantum
systems in a quite similar way. In the quantum context, we write the
Hamiltonian operator in the form
\begin{equation}\label{hq}
\mathcal{H}=\mathcal H_x+ \mathcal H_y,
\qquad \mathcal H_x= P_x^2+ V_x(X),
\qquad \mathcal H_y=P_y^2+V_y(Y)\,,
\end{equation}
where $P_x , P_y$ and $X, Y$ are the momentum and position operators, satisfying the well known commutation relations
\begin{equation}
[X,P_x ]=i \hbar\,,\qquad
[Y,P_y ]=i \hbar\, .
\end{equation}
We will work in the coordinate representation where the action of the momentum operators is given by $P_x =-i\hbar\partial_x$, $P_y =-i\hbar\partial_y$, and the action of the position operators $X, Y$ is just the multiplication by the variables $x$ and $y$, respectively. This two dimensional Hamiltonian operator \eqref{hq} can be considered integrable in the sense that it  has already two independent symmetry operators in involution:
$\mathcal H$ itself and (for example) $\mathcal A = \mathcal H_x - \mathcal H_y$. In order to get
quantum superintegrable systems of Heisenberg type, as in the classical case, we will look for
a symmetry operator,  polynomial of degree $n$ in the momentum operators
$P_x$, $P_y$, having the separated form
\begin{equation}\label{bq}
\mathcal{B}_n(X,Y,  P_x,P_y)=
\mathcal B_{nx}(X,P_x) - \mathcal B_{ny}(Y,P_y),
\end{equation}
where
\begin{equation}\label{bqx}
\mathcal B_{nx}(X,P_x)=\sum_{j=0}^n {f_j(X)\, P_x^{j}},
\qquad \mathcal B_{ny}(Y,P_y)=\sum_{j=0}^n {g_j(Y)\,P_y^{j}}\,.
\end{equation}
We will ask the component operators to satisfy Heisenberg-type
commutation relations,
\begin{equation}\label{bhqxy}
[\mathcal H_x,\mathcal B_{nx}(X,P_x)]=
[\mathcal H_y,\mathcal B_{ny}(Y,P_y)]=i\, \hbar\,,
\end{equation}
so that the symmetry condition
\begin{equation}\label{bhq}
[\mathcal H,\mathcal B_n]=  0\,
\end{equation}
is automatically satisfied. This symmetry is of order
$n$, and when $n\geq 3$ it is said to be of `higher order'. 

For a wide discussion of general third and fourth-order integrals of motion, the reader is referred to the excellent review \cite{MillerPostWinternitz2013}. 
The one dimensional case of higher order symmetries has been studied in \cite{Nikitin1997, Doebner1999}.
It is also worth to mention references
\cite{Gravel2004, Marquette2011, MillerPostWinternitz2015,MarquetteWinternitz2007,MarquetteWinternitz2008a}, dealing with higher order symmetries, which are more related to our approach.  
In the  Conclusions we will
comment on the connection between the methods and results of these references and  those obtained in the present paper.

\section{Heisenberg-type higher order integrals of motion: the classical problem}

In this Section we will investigate the existence of classical potentials and integrals of motion satisfying  Eq.~\eqref{bhcxy} and we will try to determine their explicit expressions. Here, we need to consider only one pair of the variables (for instance $x,p_x$), because the results are the same for the other variable pair. Also, in order to simplify the notation, we will take
$p_x\equiv p$, $B_{nx}(x,p)\equiv B_n$, $H_x\equiv H_n$, and therefore $v_x(x)\equiv v_n(x)$.

Notice that the PB relation (\ref{bhcxy}) can be interpreted as follows.
We can think of the Heisenberg function $-B_n$ and the Hamiltonian $H_n$
as new canonical variables $\tilde x$, $\tilde p$:
\begin{equation}\label{tilde}
\tilde H_n \equiv \tilde p = H_n(x,p),\qquad \tilde x = -
B_n(x,p)\,.
\end{equation}
The new momentum $\tilde p$ is also identified with the new Hamiltonian
$\tilde H_n$. This means that the new pair of canonical variables
$\tilde x, \tilde p$ corresponds to the characteristic function of Hamilton-Jacobi
theory \cite{Goldstein80}. We can solve the motion for the new variables:
\begin{equation}
\begin{array}{ll}
\displaystyle \dot{\tilde x} = \frac{\partial \tilde H}{\partial \tilde p} = 1,\quad
& \tilde x = t + \alpha,
\\[2.ex]
\displaystyle \dot{\tilde p} = -\frac{\partial \tilde H}{\partial \tilde x} = 0,\quad
& \tilde p = \beta,
\end{array}
\end{equation}
where $\alpha, \beta$ are constants fixed by the initial conditions.
From the motion of $\tilde x, \tilde p$, we can find the evolution
of the initial variables $x,p$ algebraically by reverting the relations (\ref{tilde}).
In summary, the problem of finding a Heisenberg system characterized
by the function $B_n$ and Hamiltonian $H_n$ is equivalent to the search of
systems such that the canonical variables of the characteristic function include
the coordinate $\tilde x$ given by a polynomial function of degree $n$ in the momentum $p$.

We start this section with a list of particular cases for some
values of $n$  in order to see some features of the potentials
$v_n(x)$ and the functions $B_n$. Later on, closed formulas for the
general $n$--order case are supplied. Finally, it is explained how
the superintegrable systems are obtained together with their
properties from the previous results.

\subsection{Particular cases}

\begin{itemize}
\item Case $n=1$

The $x$-Hamiltonian and the $x$-part of the integral of motion have the form
\begin{equation}\label{hcx1}
H_1=p^2+v_1(x),\qquad B_1=f_0(x)+ f_1(x)\,p ,
\end{equation}
and they must satisfy
\begin{equation}\label{bhcx1}
\{{H_1},B_1\}=1 .
\end{equation}
Substituting (\ref{hcx1}) in (\ref{bhcx1}) we get a set of differential equations from the coefficients of the powers $p^j,\,j=0,1,2$:
\begin{equation}
\label{bhcx1e}
 2f_1'  =  0,  \quad
 2f_0'     =   0, \quad
 1  =  f_1 \,{v'_1} , 
\end{equation}
where the prime  denotes the derivative with respect to the
argument. For the sake of simplicity, from now on we omit the
explicit dependence of the functions $f_j$ and ${v_1(x)}$ on the
variable $x$. Thus, for three unknown functions $f_0 ,\,f_1 $ and
${v_1}(x)$, there are three equations given by  (\ref{bhcx1e}). The
first two equations give $f_1 =k_1$ and $f_0 =k_0$, where $k_1$ and
$k_0$ are integration constants.
Thus, from the last equation the potential is
\begin{equation}\label{bhcx1v}
v_1(x)=\frac{x}{k_1}+c_1
\end{equation}
where $c_1$ is an irrelevant integration constant. Hence, $B_1$ takes the form:
\begin{equation}\label{cbx1}
B_1=k_0+ k_1\,p\,.
\end{equation}

\item {Case} $n=2$

For this case the  function $B_2$ is quadratic
\begin{equation}\label{hcx2}
 B_2=f_0+f_1 \,p+ f_2 \,p^2,
\end{equation}
and together with $H_2$ satisfy
\begin{equation}\label{bhcx2}
\{H_2,B_2\}=1\,.
\end{equation}
Using the Hamiltonian function and (\ref{hcx2}) in (\ref{bhcx2}), we obtain a set of equations
\begin{equation}\label{sistemaclasicoden2}
 2f_2'     =  0, \quad
 2f_1'     =  0, \quad
 2\,f_0'    =   2\,f_2 \,{v'_2}, \quad
 1  =  f_1 \,{v'_2} \,.
\end{equation}
This case lead us to the same potential:
\begin{equation}\label{bhcx2v}
v_2(x)=v_1(x)=\frac{x}{k_1}+c_1\,.
\end{equation}
The Heisenberg function $B_2$ takes the form
\begin{equation}\label{cbx2}
B_2=k_0 +k_1\,p+k_2\, v_2+ k_2\,p^2
=   k_0  + k_1 \, p +k_2 H_2= B_1 + k_2 H_2 \,.
\end{equation}
Thus,  $B_2$ is the same as $B_1$, except for a `trivial
term' proportional to the Hamiltonian corresponding to the constant $k_2$.

\item {Case} $n=3$

For this case the  function $B_3$ is cubic
\begin{equation}\label{hcx33}
 B_3=f_0+f_1 \,p+ f_2 \,p^2+ f_3 \,p^3.
\end{equation}
After imposing the condition $\{H_3,B_3\}=1$,
the coefficients $f_j$ in \eqref{hcx33} are the solutions of the set of differential equations
\begin{eqnarray}\label{sistemaclasicoden3}
 2f_3'    \!\!&\!\!=\!\!&\!\! 0, \nonumber \\
 2f_2'    \!\!&\!\!=\!\!&\!\! 0, \nonumber \\
 2\,f_1'   \!\!&\!\!=\!\!&\!\!  3\,f_3 \,{v'_3} , \\
 2\,f_0'   \!\!&\!\!=\!\!&\!\!  2\,f_2 \,{v'_3} , \nonumber \\
 1 \   \!\!&\!\!=\!\!&\  f_1 \,{v'_3} \,. \nonumber
\end{eqnarray}
If we solve  this system for the functions $f_j$, we get the following differential equation for potential
\begin{equation}\label{bhcx3ev}
\left(k_1+\frac{3}{2}\,k_3\,{v_3} \right){v'_3} =1
\end{equation}
which gives
\begin{equation}\label{bhcx3evb}
k_1 {v_3}  + \frac{3}{4}\,k_3\,{v_3}^2= x + {c_3} \,.
\end{equation}
The solution of this quadratic equation can be given explicitly,
\begin{equation}\label{bhcx3v}
v_3(x) =\frac{1}{3\,k_3}
\left(-2\,k_1\pm{2}\sqrt{3\,k_3\,x+k_1^2+3\,k_3\,c_3}\right) \,,
\end{equation}
where $c_3$ is an integration constant. The new information for this
case is obtained taking $k_1=0$ in \eqref{bhcx3v}. For instance, if
we choose $k_1 = 0$, $k_3= \pm 4/3$, $c_3= 0$ we will have the
particular solutions: \be\label{classicalvconn=3k=0} v_3(x) = \pm
\sqrt{\pm x}. \ee Remark that depending on the sign, this potential
makes sense either for $x\geq 0$ or for $x\leq 0$.

The expression of $B_3$ is
\begin{eqnarray}\label{cbx3}
B_3 \!\!&\!\!=\!\!&\!\! k_0+k_2\, v_3 +\left(k_1+\frac{3}{2}k_3\,v_3
\right)\,p+k_2\,p^2+ k_3\,p^3\nonumber
\\ 
\!\!&\!\!=\!\!&\!\! 
B_1+  k_2 H_3 + k_3 (p^3 +\frac32 v_3\, p)
= k_0+ k_1 p + k_2H_3 + k_3\frac12(p^3 - 3H_3 p).
\end{eqnarray}
Therefore, we notice that the potential $v_3(x)$ depends on three
constants: $k_1$, $k_3$ and $c_3$, but the last constant is
irrelevant since it can be eliminated by a translation in $x$. The
integral of motion $B_3$ depends on the corresponding constants,
$k_1$ and $k_3$; it also includes one additional term proportional
to the Hamiltonian, $k_2 H_3$, but this can be eliminated without
any consequence.

\item {Case} $n=4$

This leads us to the same potential as in the previous case $n=3$; $v_3(x)=v_4(x)$. But, the function $B_4$ is slightly different:
\begin{eqnarray}\nonumber
B_4\!\!&\!\!=\!\!&\!\!k_0+ k_2\, v_{4} +k_4\, v_{4} ^2+
\left(k_1+\frac{3}{2}k_3\,v_{4} \right) \,p +\left(k_2+2\,k_4\,v_{4}
\right)\,p^2+k_3\,p^3+k_4\,p^4
\\[1.5ex]
\!\!&\!\!=\!\!&\!\! B_1+k_2\, H_4 + k_3(p^3 +\frac32 v_4 p) + k_4\,
(H_4)^2 .\label{cbx4}
\end{eqnarray}

\item {Case} $n=5$

Now $B_5$ is a fifth-order polynomial in $p$
\begin{equation}\label{hcx55}
 B_5=f_0+f_1 \,p+ f_2 \,p^2+ f_3 \,p^3+ f_4 \,p^4+ f_5 \,p^5.
\end{equation}
Imposing the condition $\{H_5,B_5\}=1$,
the functions $f_j$ in \eqref{hcx55} turn out to be the solutions of the set of differential equations
\begin{eqnarray}\label{sistemaclasicoden5}
 2f_5'    \!\!&\!\!=\!\!&\!\! 0, \nonumber \\
 2f_4'    \!\!&\!\!=\!\!&\!\! 0, \nonumber \\
 2\,f_3'   \!\!&\!\!=\!\!&\!\!  5\,f_5 \,{v'_5} ,  \nonumber \\
 2\,f_2'   \!\!&\!\!=\!\!&\!\!  4\,f_4 \,{v'_5} , \\
 2\,f_1'   \!\!&\!\!=\!\!&\!\!  3\,f_3 \,{v'_5} , \nonumber \\
 2\,f_0'   \!\!&\!\!=\!\!&\!\!  2\,f_2 \,{v'_5} , \nonumber \\
 1 \   \!\!&\!\!=\!\!&\ f_1 \,{v'_5} \,. \nonumber
\end{eqnarray}
After solving this system for the functions $f_j$,
the equation for the potential has the form
\begin{equation}\label{bhcx5ev}
\left(k_1+\frac{3}{2}\,k_3\,{v_5} +\frac{15}{8}\,k_5\,{v_5}^2\right){v'_5} =1
\end{equation}
or
\begin{equation}\label{bhcx5evb}
k_1{v_5} +\frac{3}{4}\,k_3\,{v_5}^2+\frac{5}{8}\,k_5\,{v_5}^3=x+ c_5
\end{equation}
from which the solution can be explicitly obtained. For instance, if
we concentrate on the particular values $k_1=k_3=c_1=0$, $k_5 =
8/5$, we get
 \begin{equation}
\label{n=5k_1=k_3=0vvv5}
v_5 (x) =  \sqrt[3]{ x  }.
\end{equation}
For this case, the Heisenberg function $B_5$ is
\begin{eqnarray} \nonumber
B_5 \!\!&\!\!=\!\!&\!\!k_0+k_4\,
v_5+\left(k_1+\dfrac{3}{2}k_3\,v_5+\dfrac{15}{8}k_5\,v_5^2\right)\,p+\left(
k_2+2\,k_4\,v_5
\right)\,p^2+\left(k_3+\dfrac{5}{2}k_5\,v_5\right)\,p^3 \\
[1.5ex] \!\!&\!\! \!\!&\!\!   +k_4\,p^4+k_5\,p^5 \, \nonumber
\\[1.5ex]
\!\!&\!\!=\!\!&\!\! k_0+k_1 p +k_2\, H_5 +k_3(p^3 +\frac32 v_5 p)
+k_4\, (H_5)^2 + k_5(p^5 + \frac52 v_5 p^3 +\frac{15}{8} v_5^2 p)  .
\label{cbx5}
\end{eqnarray}
We see that the potential $v_5(x)$ given in equation
(\ref{bhcx5evb}) depends on the constants $k_1$, $k_3$ and $k_5$
(the constant $c_5$ can be eliminated as before by means of a
translation). With respect to the function $B_5$ it depends on these
three constants (the terms including $k_4, k_2$ and $k_0$ can be
omitted). The coefficients of these constants have the same
expressions in terms of $v(x)$ as the corresponding ones in the
previous cases, except for the new one corresponding to $k_5$:
\begin{equation}\label{b50c}
k_5 B_5^0 \equiv k_5(p^5 + \frac52 v_5 p^3 +\frac{15}{8} v_5^2 p) = k_5 \frac38
(p^5 - \frac{10}{3} p^3 H_5 + 5 p H_5^2)\, .
\end{equation}
\item {Case} $n=6$

The  case $n=6$, gives us the same equation for potential and
therefore the same potential as in the previous case $n=5$: $v_5(x)=v_6(x)$.
The function $B_6$ differs from $B_5$ in a trivial term proportional
to  $H_6^3$.


\item {Case} $n=7$

The integral of motion will be
\begin{equation}\label{hcx77}
 B_7=f_0+f_1 \,p+ f_2 \,p^2+ f_3 \,p^3+ f_4 \,p^4+ f_5 \,p^5,
\end{equation}the coefficients satisfying the system
\begin{eqnarray}\label{sistemaclasicoden77}
 2f_7'    \!\!&\!\!=\!\!&\!\! 0, \nonumber \\
 2f_6'    \!\!&\!\!=\!\!&\!\! 0, \nonumber \\
 2\,f_5'   \!\!&\!\!=\!\!&\!\!  7\,f_7 \,v'_7 ,  \nonumber \\
 2\,f_4'   \!\!&\!\!=\!\!&\!\!  6\,f_6 \,v'_7 ,  \nonumber \\
 2\,f_3'   \!\!&\!\!=\!\!&\!\!  5\,f_5 \,v'_7  ,   \\
 2\,f_2'   \!\!&\!\!=\!\!&\!\!  4\,f_4 \,v'_7  , \nonumber \\
 2\,f_1'   \!\!&\!\!=\!\!&\!\!  3\,f_3 \,v'_7 , \nonumber \\
 2\,f_0'   \!\!&\!\!=\!\!&\!\!  2\,f_2 \,v'_7 , \nonumber \\
 1\    \!\!&\!\!=\!\!&\  f_1 \,v'_7  \,. \nonumber
\end{eqnarray}
The solution of these equations lead to
\begin{equation}\label{bhcx77ev}
\left(k_1+\frac{3!!}{2}\,k_3\, v_7 +\frac{5!!}{2^2}\,k_5\,\frac{v_7^2}{2!} +\frac{7!!}{2^3}\,k_7\frac{v_7^3}{3!}
\right) v'_7 =1
\end{equation}
or
\begin{equation}\label{bhcx77evb}
k_1 v_7 +\frac{3!!}{2}\,k_3\frac{v_7^2}{2!} +\frac{5!!}{2^2}\,k_5\,\frac{v_7^3}{3!} +\frac{7!!}{2^3}\,k_7\frac{v_7^4}{4!}=x+ c_7,
\end{equation}
which is an implicit expression of the potential. The new relevant
information in this equation is obtained for $k_7\neq 0$. In
particular if $k_1=k_3=k_5=c_7=0$ and $k_7 =\pm \frac{2^3 4!}{7!!}$,
we have the potential: \be \label{classicalvconn=7k=0} v_7 (x)= \pm
\sqrt[4]{\pm x}. \ee
Remark that depending on the sign of $k_7$
this potential makes sense either for $x\geq 0$ or for $x\leq 0$.

\end{itemize}


\subsection{General case}

The functions $B_n$ found above can be expressed in two ways:
(a) collecting the terms by the integration constants $k_j$,
or (b) grouping terms in powers of $p$.

\medskip
\noindent
{\it (a) Solutions in terms of $k_i$}

We have seen that the problems for consecutive odd and even degrees
$n=2\ell +1$ and $n=2\ell +2$ have essentially the same solutions.
Therefore, let us restrict to an odd function $B_{n}$, $n=2\ell +1$,
$\ell = 0,1,\dots$, and the corresponding potential
$v_{2\ell+1}(x)$, then their expressions take the general form
\be\label{solks}
 B_{2\ell +1} = { \sum_{j=0}^{\ell} k_{2j+1} b_{2j+1}(p)},
\qquad
 \sum_{j=0}^{j=\ell}  k_{2j+1} \alpha_{2j+1}{v_{2\ell+1}^{j+1}} = x + {c_{2\ell+1}},
\ee
where $b_{2j+1}(p)$ are polynomials of degree $2j+1$ of $p$ and
$\alpha_{2j+1}$ are constants which are obtained in the integration process,
while $k_{2j+1}$ and $c_{2\ell+1}$ are arbitrary integration constants.
Thus, only the odd integration constants are important, the even ones do not
play any role in the solutions.

There are two particular cases worth to mention.
\begin{itemize}
\item[(i)]
If $k_{2\ell+1}=0$, but ${k_{j}} \neq 0$, $\forall j<\ell$, then the formulas
(\ref{solks}) valid for $n= 2\ell +1$ come into the ones for the previous case
$n= 2(\ell-1) +1$. In other words, the formulas for $n = 2\ell +1$, include
as particular cases all the formulas for the previous cases.

\item[(ii)]
If $k_{2\ell+1}\neq 0$, but ${k_{j}} = 0$, $\forall j<\ell$, then we get that
the potential is a root,
\begin{equation}
{v_{2\ell +1}}(x)= (a\, x +b)^{ 1/(\ell+1)},
\end{equation}
where $a,b$ depend on the integration constants, while ${B_{2\ell+1}
= k_{2\ell +1} b_{2\ell +1}(p)}$ is a polynomial of degree $2\ell
+1$, according to (\ref{solks}). In conclusion, we can say that this
type of superintegrable potentials include all the roots
$v_{2\ell+1}(x)\propto x^{1/(\ell+1)}$, starting with the trivial
linear potential $v_1(x) \propto x$.
\end{itemize}

\medskip
\noindent
{\it (b) Solutions in powers of $p$}

Now, we will deal with the general case of the polynomial function expressed
in powers of $p$ as
\begin{equation}
B_{n}(x,p)=\sum_{j=0}^{n}f_{j}(x)\,p^j
\end{equation}
and substitute it in the PB equation (\ref{bhcxy}). Then we get a list of equations for the
coefficients $f_j$ and the potential $v(x)$,
\begin{eqnarray}\label{efes}
2 f'_n   \!\!&\!\!=\!\!&\!\!  0, \nonumber \\
2 f'_{n-1}   \!\!&\!\!=\!\!&\!\!  0, \nonumber \\
2 f'_{n-2}  \!\!&\!\!=\!\!&\!\!  n f_n v' , \nonumber  \\
2 f'_{n-3}   \!\!&\!\!=\!\!&\!\!  (n-1) f_{n-1} v' , \nonumber \\
2 f'_{n-4}   \!\!&\!\!=\!\!&\!\!  (n-2) f_{n-2} v', \\
 \vdots  \quad   \!\!&\!\!\!\!&\!\! \quad \vdots  \nonumber\\
2 f'_{0}  \!\!&\!\!=\!\!&\!\!  2 f_2 v'  , \nonumber \\
  1 \ \  \!\!&\!\!=\!\!&\!\!  f_1 v'. \nonumber
\end{eqnarray}
{Remark that for every value of $n$ this system can be completely
separated into two: one only for the odd-index ($f_1,f_3,\dots$)
functions and another only for the even-index functions
($f_0.f_2,\dots$).}

We introduce the following notation
\begin{equation}
B_{n}(x,p)=f_{\frac{n}{2},0}\,p^{n}+f_{\frac{n-1}{2},0}\,p^{n-1}+f_{\frac{n}{2},1}\,p^{n-2}+f_{\frac{n-1}{2},1}\,p^{n-3}+\dots+
\left\{\begin{array}{ll}
f_{\frac{n}2,\frac{n}2}\,, \ &{\rm for}\ n \ {\rm even}
\\[1.5ex]
f_{\frac{n-1}2,\frac{n-1}2}\,, \ &{\rm for}\ n \ {\rm odd}
\end{array}\right.
\end{equation}
where the `old'  $f_j$ and `new' $f_{\mu,\nu}$ coefficients  are related as follows:
\begin{equation}
f_n=f_{\frac{n}{2},0},
\qquad f_{n-1}=f_{\frac{n-1}{2},0},
\qquad f_{n-2}=f_{\frac{n}{2},1},
\qquad f_{n-3}=f_{\frac{n-1}{2},1},\,\dots
\end{equation}
Then, equation (\ref{efes}) can be integrated and
the coefficients $f_{\mu,\nu}$ are given by
\begin{equation}\label{fgeneral}
f_{\mu,\nu} = \sum_{\nu'=0}^{\nu} \frac{\Gamma(\mu + 1 -\nu')}{\Gamma(\mu +1 -\nu)}
\, \frac{v^{\nu - \nu'}}{(\nu -\nu')!} \, k_{2(\mu - \nu')}\,.
\end{equation}
As only the odd index cases seem to be relevant, let us assume that
${n=2\ell+1}$ is odd. Then if we substitute the formula for $f_1$
given by the last equation of  (\ref{efes}), we will get
\begin{equation}\label{fgeneral1}
f_1\, v' \equiv f_{{\ell +\frac{1}2,\ell}}\, v'
=\left(\sum_{\nu'=0}^{{\ell}} \frac{\Gamma({\ell+3/2}
-\nu')}{\Gamma(3/2)} \, \frac{v^{{\ell} - \nu'}}{({\ell}  -\nu')!}
\, k_{{ 2\ell +1 - 2\nu'}}\right)v' = 1\,.
\end{equation}
Thus, we get the algebraic equation for the potential:
\begin{equation}\label{fgeneral2}
\sum_{\nu'=0}^{{\ell}} \frac{\Gamma({\ell+3/2}  -\nu')}{\Gamma(3/2)}
\, \frac{v^{{\ell}+1 - \nu'}}{({\ell} +1 -\nu')!} \, k_{{ 2\ell +1 -
2\nu'}} = x+ c_n\,.
\end{equation}

\subsection{Superintegrable Hamiltonians and integrals of motion}

In this Subsection we will discuss  different ways to construct superintegrable
Hamiltonians from the previous results.
\medskip

\noindent
{\it a) Superintegrable Hamiltonians by adding two Heisenberg Hamiltonians}

Once we have the Heisenberg algebras $\langle H_{nx}, B_{nx}, 1\rangle$,
$\langle H_{ny}, B_{ny}, 1\rangle$ we can write the superintegrable Hamiltonian
\begin{equation}
H_{n} = H_{nx} + H_{ny}, \qquad v_{nn}(x,y) = v_{nx}(x) + v_{ny}(y),
\end{equation}
whose `extra' constant of motion of odd degree $n=2\ell+1$ is given
by
\begin{equation}
B_{n} = B_{nx} - B_{ny}\,.
\end{equation}
Remark that, in general these systems are only defined in a region of the plane
${\mathbb R}^2$. Some examples are:
\begin{itemize}
\item[(i)] $v_{11}(x,y) = \alpha\, x + \beta\, y$.

\item[(ii)] $v_{33}(x,y) = \alpha\, \sqrt{x} + \beta\, \sqrt{y}$,\quad  $x\geq 0, y\geq 0$.

\item[(iii)] $v_{55}(x,y) = \alpha\, x^{1/3} + \beta\,  y^{1/3}$.

\item[(iv)]  $v_{77}(x,y) = \alpha\, x^{1/4} + \beta\,  y^{1/4}$,\quad  $x\geq 0, y\geq 0$.

\end{itemize}
Example (ii) was considered in \cite{MarquetteWinternitz2008a} as case 5.

In fact, we can build superintegrable systems by combining Heisenberg algebras of different orders,
\begin{equation}
H_{mn} = H_{mx} + H_{ny}, \qquad v_{mn}(x,y) = v_{mx}(x) +
v_{ny}(y),
\end{equation}
whose `extra' constant of motion of odd degree ${\rm
max}(m=2k+1,n=2\ell+1)$ will be given by
\begin{equation}
 B_{2k+1,2\ell+1} = B_{2k+1\, x} - B_{2\ell+1\, y}\,.
\end{equation}
The corresponding potentials include  linear combinations of
different roots, \be v_{2k+1,2\ell+1}(x,y) = \alpha\, x^{1/(k+1)} +
\beta\,  y^{1/(\ell+1)}. \ee
Some special cases are:
\begin{itemize}
\item[(v)] $v_{13}(x,y) = \alpha\, x + \beta\, \sqrt{y}$,\quad  $y\geq 0$.

\item[(vi)] $v_{35}(x,y) = \alpha\, \sqrt{x} + \beta\,  y^{1/3}$,\quad  $x\geq 0$.

\item[(vii)] $v_{15}(x,y) = \alpha\, x  + \beta\,  y^{1/3}$.

\end{itemize}
Example (v) was included in \cite{MarquetteWinternitz2008a} as case
7. Notice that since each one dimensional potential  $v_{mx}(x)$ or
$v_{ny}(y)$ is a monotonous function, these potentials $v_{nm}$ will
not allow for any bounded motion.

\medskip

\noindent
{\it b) Global and local Hamiltonians in $\mathbb R^2$}

\begin{figure}
\centering
\includegraphics[width=0.5\textwidth]{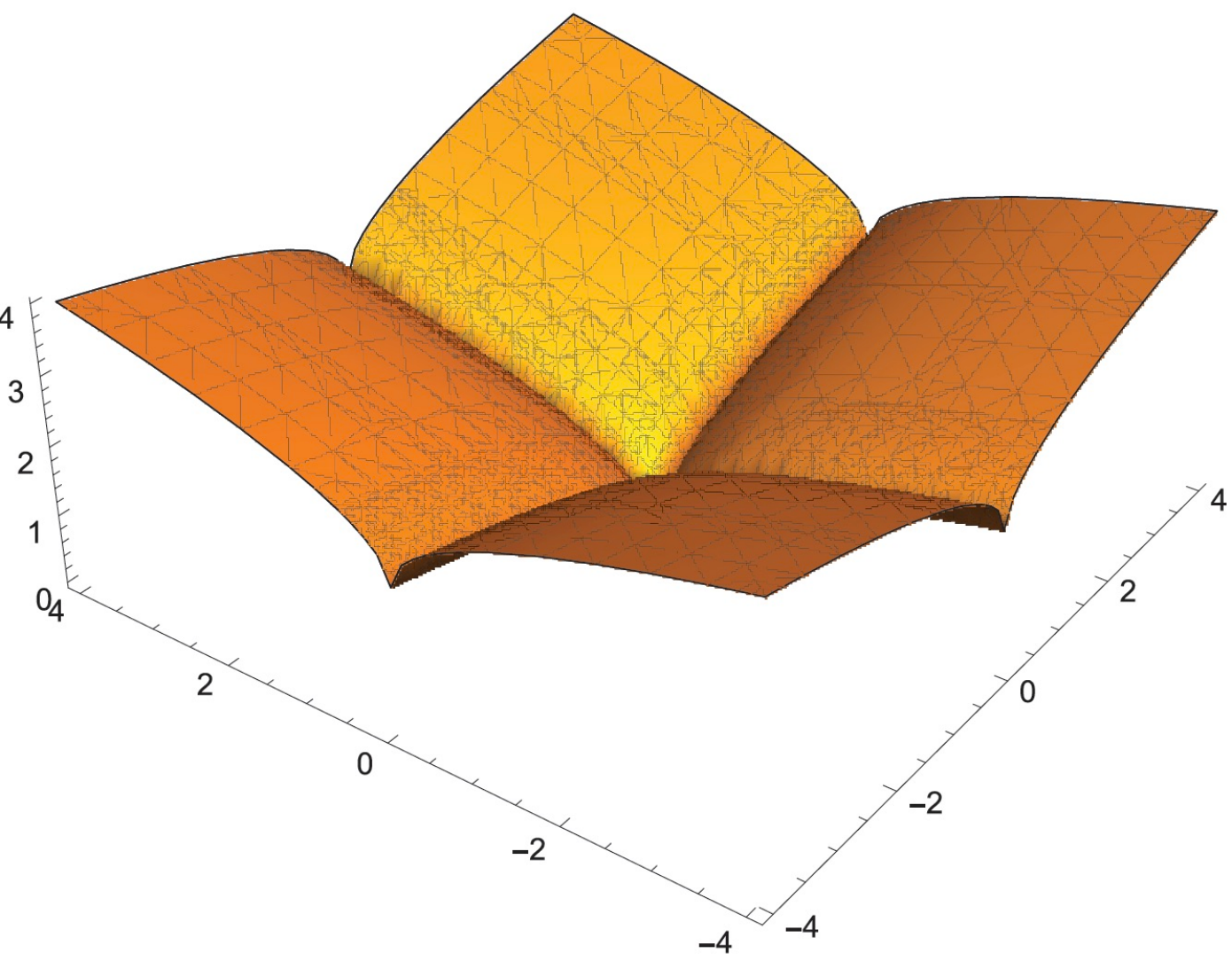}
\caption{\small Plot of the potential
$\tilde v_{33} = \sqrt{|x|} + \sqrt{|y|}$. \label{fig1}}
\end{figure}

\begin{figure}
\centering
\includegraphics[width=0.4\textwidth]{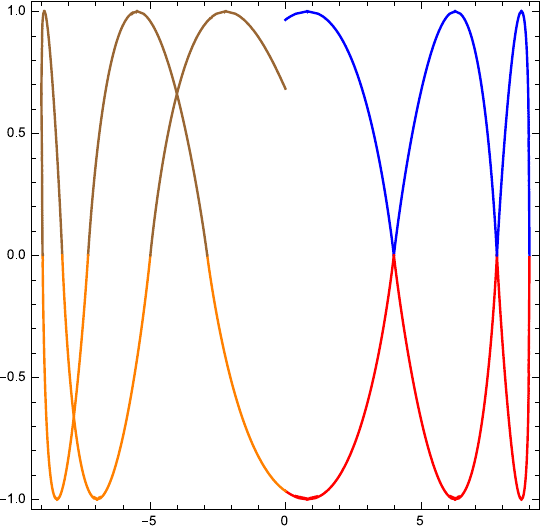}
\qquad\quad
\includegraphics[width=0.4\textwidth]{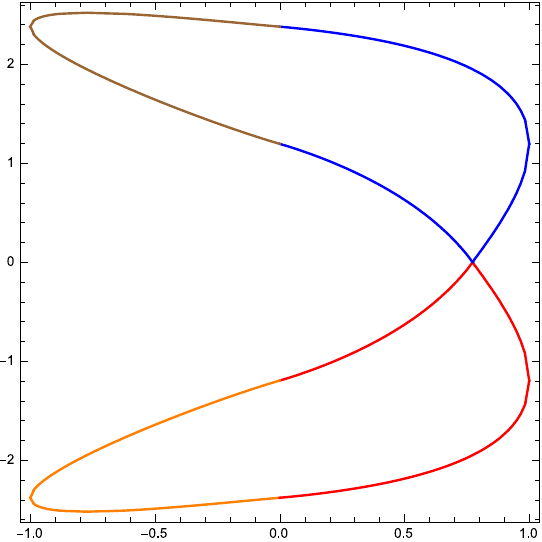}
\caption{\small Plot of a non periodic trajectory corresponding to
parameters $E_x=3, E_y=1, B=5$ (left) and a closed periodic trajectory
with parameters  $E_x=1, E_y=\sqrt[3]{4}, B=1$ (right).
The pieces of trajectory corresponding
to each quadrant of the plane are in a different color.\label{fig2}}
\end{figure}

As it is clear from example (ii) given above, some of the
superintegrable Hamiltonians have potentials $v_{mn}(x,y)$ defined
only in a region ${\cal D}$ of the plane. We may try to extend this
Hamiltonian to the whole plane by pasting it with other
superintegrable systems with potentials $v_{mn}^i(x,y)$ defined in
disjoint (except for their boundaries) regions  ${\cal D}^i$, such
that they cover the whole plane, $\cup_i {\cal D}^i = \mathbb R^2$.
For instance,  considering the particular case $\alpha=\beta=1$, we
can `complete' the potential of case (ii) as follows:
\begin{equation}\label{vt33}
\tilde v_{33}(x,y) =\sqrt{|x|} + \sqrt{|y|}=\left\{ \begin{array}{ll}
v_{33}^1 = \sqrt{x} + \sqrt{y},\quad &{\cal D}^1 = \{ (x,y), x\geq 0, y\geq 0 \}
\\[1.5ex]
v_{33}^2=\sqrt{-x} + \sqrt{y},\quad &{\cal D}^2 = \{ (x,y),x\leq 0, y\geq 0 \}
\\[1.5ex]
v_{33}^3=\sqrt{-x} + \sqrt{-y},\quad &{\cal D}^3 = \{ (x,y),x\leq 0, y \leq 0 \}
\\[1.5ex]
v_{33}^4=\sqrt{x} + \sqrt{-y},\quad &{\cal D}^4 = \{ (x,y),x\geq 0, y\leq 0 \}
\end{array}\right.
\end{equation}
Another  example extended from case (i) is given by
\begin{equation}\label{vt11}
\tilde v_{11} = |x| + |y|\,.
\end{equation}
We can apply this `pasting process' in order to produce a global
potential such that it will allow for bounded trajectories, as it is
the case of the extensions (\ref{vt33}) or (\ref{vt11}), given
above. However, in these cases, the resulting system with global
potential $\tilde v_{33}(x,y)$ will not be superintegrable since
there is not a  `global' constant of motion for $\tilde v$. In a
motion under $\tilde v_{33}(x,y)$, when the particle is in ${\cal
D}^1$ the constant of motion is $B^1$, but when the particle crosses
from the domain ${\cal D}^1$ to ${\cal D}^2$, the constant of motion
will change and it will take a different value $B^2$, and so on. In
this way after the particle has been $n_1$ times in the region
${\cal D}^1$, the constant of motion $B^1$ will have taken, in
general, $n_1$ different values. Therefore, in general the motion
will not be periodic. In Fig. \ref{fig1} it is shown the plot of the
`pasted potential' given by (\ref{vt33}). The motion of a particle
in this potential is a superposition of two one-dimensional motions
corresponding to the potentials $\tilde v_x = \sqrt{|x|}$ and
$\tilde v_y = \sqrt{|y|}$. In this case, the ratio of the
frequencies $\nu_x$ and $\nu_y$ is obtained by action-angle
variables method and it is given by
\begin{equation}\label{aa}
\frac{\nu_x}{\nu_y} = \left(\frac{E_x}{E_y}\right)^{3/2}\, .
\end{equation}
Only when this ratio is a rational number, the bounded motion in the plane
will be periodic. Examples of periodic and non-periodic orbits for this
pasted potential are given in
Fig. \ref{fig2}.
For `global' superintegrable systems all the bounded motions are periodic
and the trajectories look like deformed Lissajous curves \cite{CKN2014,CKNO2013}.
We should remark that such trajectories are smooth and they do not present `angles'
when crossing the boundary of a domain. This point was not clearly explained in previous references \cite{MarquetteWinternitz2007,MarquetteWinternitz2008a}.

\medskip

\noindent
{\it c) Superintegrable Hamiltonians in higher dimensions}

We can define superintegrable Hamiltonians in three or more dimensions. For instance,
we simply add three one dimensional Heisenberg Hamiltonians  to get a three dimensional
Hamiltonian
\begin{equation}
H_{nmp}=H_{nx}(x)+ H_{my}(y)+H_{pz}(z)\,.
\end{equation}
This new Hamiltonian is superintegrable. First of all, as it is separable,
we have three constants of motion in involution, for instance: $H_{nx}(x)$, $H_{my}(y)$
and $H_{pz}(z)$. Second, as each Hamiltonian is of Heisenberg type,
we can construct two additional independent constants of motion:
\begin{equation}
B_1 = B_{nx} - B_{my} ,\qquad B_2 =  B_{my} - B_{pz}\,.
\end{equation}
The symmetry algebra of all the independent constants of motion
is very easy to compute, it is essentially a subalgebra of the direct sum of
 one-dimensional Heisenberg algebras.

\section{Heisenberg-type higher order symmetries: the quantum problem}

In the framework of quantum mechanics, let us consider now a Hamiltonian operator
$\mathcal{H}$ in one Cartesian coordinate ($2m=1$):
\begin{equation}\label{quantum}
 \mathcal{H}=P^2+V(X),
\end{equation}
where $P$ and $X$ are the momentum and position operators, satisfying the well known commutation relation
\begin{equation}
[X,P]=i \hbar.
\end{equation}
In the sequel, we will work in the coordinate representation of wave functions where the momentum operator is
$P=-i\hbar d/dx$ and the position operator $X$ is just the multiplication by the variable $x$.

Now, we introduce the notion of Heisenberg operator $\mathcal{B}_n$
for this Hamiltonian $ \mathcal{H}$ as a $n$-order polynomial in the
momentum operator $P$ with $x$-dependent coefficients, where the
powers of the operator $P$ are placed `on the right', that is:
\begin{equation}\label{bquant}
\mathcal{B}_n=\sum_{k=0}^n {F_k(x)\, P^{k}},
\end{equation}
being $F_k(x)$ functions of the real variable $x$, to be determined. The Heisenberg function (\ref{bquant}) must satisfy the following commutation relation with the
Hamiltonian (\ref{quantum})
\begin{equation}\label{bhquant}
[\mathcal{H},\mathcal{B}_n]=i \hbar.
\end{equation}
This condition eventually will gives us a condition on the potential, that will depend on $x$, but also on $n$; due to this fact, instead of (\ref{quantum}) we will use the following notation for the Hamiltonian
\begin{equation}\label{quantum2}
 \mathcal{H}_n=P^2+V_n(x).
\end{equation}
Remark that if we take the formal adjoint of relation (\ref{bhquant}) we get
\begin{equation}\label{bhquant2}
[\mathcal{H},\mathcal{B}_n^\dagger]=i \hbar,
\end{equation}
where $\mathcal{B}_n^\dagger$ is the adjoint differential operator of
$\mathcal{B}_n$. Therefore,
\begin{equation}\label{bhquant22}
[\mathcal{H},\mathcal{B}_n^s]=i \hbar,\qquad \mathcal{B}_n^s =
\frac12(\mathcal{B}_n + \mathcal{B}_n^\dagger)\, .
\end{equation}
Hence, we can always assume the operator $\mathcal{B}_n$ in
(\ref{bhquant})  to be a Hermitian differential operator.

In the following, for the sake of simplicity we will omit the explicit dependence of the functions $F_\ell(x)$ and $V_\ell(x)$ on the variable $x$.

\subsection{The potentials and the Heisenberg operators}

Now, we will find the form of the potentials and the operators
$\mathcal{B}_n$ satisfying equation (\ref{bhquant}), for different values of $n$.
\begin{itemize}
\item Case $n=1$

The Heisenberg operator has the form
\begin{equation}\label{n1quantb1}
\mathcal{B}_1=  F_0+ F_1 \,P,
\end{equation}
similar to its classical equivalent \eqref{hcx1}. Substituting (\ref{n1quantb1}) in (\ref{bhquant}), with $P=-i\hbar d/dx$, we get the following system of ordinary differential equations
\begin{eqnarray}\label{efesquantn=1}
 F_1'   = 0, \qquad
 F_0'   = 0   , \qquad
 F_1  V'_1  =  1.
\end{eqnarray}
where the primes denotes the derivative with respect to the variable
$x$. Observe that this system is exactly the same obtained in the
classical situation \eqref{bhcx1e}, without any `quantum
corrections'. Therefore, it is trivially solved and we get the same
solutions as in the classical case:
\begin{equation}\label{quantpotn=1}
F_1 =k_1, \quad  F_0 =k_0, \quad V_1(x)=\frac{x}{k_1}, \quad \mathcal{H}_1=P^2+\frac{x}{k_1},\quad \mathcal{B}_1=k_0+k_1\, P,
\end{equation}
where $k_1$ and $k_0$ are integration constants;  we have omitted a third irrelevant additive integration constant in the potential.

\item Case $n=2$

The Heisenberg operator takes the form
\begin{equation}\label{n2quantb2}
\mathcal{B}_2=  F_0 + F_1 \,P+ F_2 \,P^2.
\end{equation}
Substituting (\ref{n2quantb2}) in (\ref{bhquant}) we get the following set of equations:
\begin{eqnarray}\label{efesquantn=2}
 F_2'    = 0, \qquad
 F_1'   = 0 , \qquad
 F_0'= F_2  V_2' ,\qquad
1 =F_1 V_2' +  i\hbar  (F_0''  -   F_2  V_2'' ).
\end{eqnarray}
Here we have  a quantum version of the classical result \eqref{sistemaclasicoden2}, with a quantum correction of first order in $\hbar$. Nevertheless, this $\hbar$ dependence is only apparent because the coefficient of $\hbar$ turns out to be zero. In fact, the solution of this system is:
\begin{equation}
F_2 =k_2, \qquad F_1 =k_1, \qquad  F_0 =k_0+k_2 V_2 ,
\end{equation}
where $k_0, \, k_1$ and $k_2$ are integration constants ($k_0$ is irrelevant and will be neglected), and the potential is exactly the same obtained in the case $n=1$ given in equation (\ref{quantpotn=1}):
\begin{equation}\label{quantpotn=2}
V_2(x)=\frac{x}{k_1}=V_1(x), \qquad \mathcal{H}_2=P^2+\frac{x}{k_1}.
\end{equation}
Hence, the operator $\mathcal{B}_2$ has the expression:
\begin{equation}\label{quantbn=2final}
\mathcal{B}_2=k_1\,P+ k_2\left(P^2 + V_2 \right)
= {B}_1+ k_2  {H}_2 .
\end{equation}
The term $k_2 {H}_2$ is redundant, and therefore the nontrivial solution reduces to $k_1\,P$.
Hence, the case $n=2$ does not provide new interesting results, as its solution coincides exactly with the one obtained in case $n=1$.

\item Case $n=3$

Now, the Heisenberg operator is
\begin{equation}\label{n3quantb3}
\mathcal{B}_3=  F_0 + F_1 \,P+ F_2 \,P^2+ F_3 \,P^3.
\end{equation}
Substituting (\ref{n3quantb3}) in (\ref{bhquant}) we get the following set of equations:
\begin{eqnarray}\label{efesquantn=3}
2F_3'    \!\!&\!\!=\!\!&\!\!  0 , \nonumber \\
2 F_2'     \!\!&\!\!=\!\!&\!\!  0 , \nonumber \\
2 F_1'     \!\!&\!\!=\!\!&\!\! 3 F_3  V_3'   ,   \\
2 F_0'    \!\!&\!\!=\!\!&\!\!  2F_2  V_3' + i\hbar( F_1''   -3 F_3 V_3'' ) ,  \nonumber \\
 1 \ \,   \!\!&\!\!=\!\!& \,  F_1 V_3'  + i\hbar (F_0'' - F_2 V_3'' ) -\hbar^2 F_3 V'''_3  . \nonumber
\end{eqnarray}
Observe the presence on the right hand side of this system of some
nontrivial quantum corrections which obviously were not present in
the classical case \eqref{sistemaclasicoden3}: nonvanishing terms
containing powers of $\hbar$, that eventually will make more
difficult to find the solution. Indeed, the `quantum system' does
not have the separation property between the functions with even
indices ($F_0,F_2,\dots$) and odd indices ($F_1,F_3,\dots$),
something that is true in the classical case, as can be clearly seen
in \eqref{efes}.

The solution of the system \eqref{efesquantn=3} exhibits the explicit presence of  quantum terms:
\begin{equation}
F_3 =k_3, \quad F_2 =k_2, \quad F_1 =k_1+\frac32 k_3 V_3 , \quad F_0 =k_0+k_2 V_3  -\frac{3}{4} i \hbar {k_3}  V_3' ,
\end{equation}
where the $k_\ell$ are integration constants ($k_0$ will be omitted
in the sequel), and the potential $V_3(x)$ must satisfy the
following third order nonlinear differential equation
\begin{equation}\label{quantpotn=3}
{k_3} \left( \hbar^2 V_3'''  -6V_3 V_3'  \right) -4 {k_1}V_3'  +4=0,
\end{equation}
which is the ``quantum version'' of \eqref{bhcx3ev}: indeed, if we make $\hbar=0$ in \eqref{quantpotn=3}
we recover \eqref{bhcx3ev}.

The nonlinear differential equation  \eqref{quantpotn=3} can be integrated once:
\begin{equation}\label{painleve}
{k_3} \left( \hbar^2 V_3''  -3V^2_3  \right) -4 {k_1}V_3 +4x+c_3=0.
\end{equation}
Remark that the information really new in the case $n=3$ comes from
the term with coefficient $k_3$: indeed, if for example we put
$k_3=0$ in \eqref{painleve} we go back to the result
\eqref{quantpotn=2}. If we take $k_1=0$ we have \be {k_3} \left(
\hbar^2 V_3''  -3V^2_3  \right) +4x+c_3=0 
\ee 
that can be finally
reduced to the first Painlev\'e equation
\be\label{painleve2}
\frac{d^2 W_3}{dz^2}=6W_3^2+z \ee by means of the following
transformations: 
\be
x=-{\rm sign} (k_3)\ \sqrt[5]{\frac{|k_3| \hbar^4}{2}}\ z-\frac{c_3}4, \qquad
V_3= 2 \left(\frac{2\hbar}{|k_3|}\right)^{2/5}  \ W_3.
\ee
Notice that if we take $\hbar=0$ in  \eqref{painleve}, the equation
reduces to the classical solution obtained previously in
\eqref{classicalvconn=3k=0}, without necessity of any further
integration. Obviously in the quantum case the presence of the terms
with powers of $\hbar$ make the whole story more interesting and
also more difficult to deal with.

The operator $\mathcal{B}_3$ can be written as
\begin{equation}\label{quantbn=3final}
\mathcal{B}_3=
\mathcal{B}_1+ k_2  \mathcal{H}_3 +
k_3 \mathcal{B}_3^0,
\end{equation}
and, obviously, only the $\mathcal{B}_3^0$ term  gives us new interesting information:
\begin{equation}\label{quantbn=3zero}
\mathcal{B}_3^0= P^3 +\frac32 V_3  P -\frac{3}{4} i \hbar V_3'= P^3 +\frac32  \,\frac{V_3  P+ P V_3 }2
=-\frac12\left(  P^3 -3  \, \frac{\mathcal{H}_3 P+ P \mathcal{H}_3}2  \right).
\end{equation}
We see that $\mathcal{B}_3^0$ is the symmetrized version of
the classical expression (\ref{cbx3}).


\item Case $n=4$

The integral of motion is now of fourth order. The problem is solved
as in the previous cases and we obtain that the potential $V_4(x)$
must satisfy a third order nonlinear differential equation which is
exactly the same as  (\ref{quantpotn=3}) for $V_3(x) $, therefore
$V_4(x) =V_3(x)$. The integral of motion $\mathcal{B}_4$ has the
form:
\begin{equation}\label{quantbn=4final}
\mathcal{B}_4=\mathcal{B}_1+k_2 \mathcal{H}_4+k_3 \mathcal{B}_3^0 + k_4 (\mathcal{H}_4)^2.
\end{equation}
There is no new interesting information, because the terms containing powers of $\mathcal{H}_4$ are irrelevant,
and the other two components, $\mathcal{B}_1$ and $\mathcal{B}_3^0$ have been already obtained.
A similar situation appeared in case $n=2$, and is typical of all the even cases.


\item Case $n=5$

The integral of motion is
\begin{equation}\label{n5quantb5}
\mathcal{B}_5=  F_0 + F_1 \,P+ F_2 \,P^2+ F_3 \,P^3+ F_4 \,P^4+ F_5 \,P^5.
\end{equation}
Substituting (\ref{n5quantb5}) in (\ref{bhquant}), we get the
following set of equations:
\begin{eqnarray}\label{efesquantn=5}
2 {F_5}'  \!\!&\!\!=\!\!&\!\!  0 ,\nonumber \\
2 {F_4}'  \!\!&\!\!=\!\!&\!\!  0 , \nonumber \\
 2{F_3}'   \!\!&\!\!=\!\!&\!\! 5{F_5} V_5' , \nonumber  \\
2{F_2}'\!\!&\!\!=\!\!&\!\! 4{F_4}V_5'+ i\hbar ({F_3}'' -10 {F_5} V_5''),\\
2 {F_1}'\!\!&\!\!=\!\!&\!\! 3{F_3}V'_5 + i \hbar( {F_2}'' -6{F_4} V''_5) -10 \hbar^2 {F_5} V'''_5   , \nonumber \\
2{F_0}'\!\!&\!\!=\!\!&\!\! 2 {F_2}V'_5+ i\hbar ({F_1}''-3{F_3}V''_5)-\ 4 \hbar^2 {F_4}V'''_5 + 5 i\hbar^3 {F_5}V_5^{(iv)}  ,\nonumber  \\
 1 \ \,   \!\!&\!\!=\!\!& \, F_1 V'_5+ i \hbar ({F_0}''-  F_2 V_5'') \ \, -\ \; \hbar^2F_3V_5'''\; +\ i\hbar^3F_4V_5^{(iv)}+\hbar^4 {F_5} V_5^{(v)}. \nonumber
\end{eqnarray}
Observe that this quantum version of the classical result \eqref{sistemaclasicoden5} has quantum corrections up to order $\hbar^4$.
Looking at \eqref{efesquantn=5} it is quite obvious that the quantum corrections are growing in importance.
In spite of the imposing aspect of this system, it is possible to find the explicit solution of the functions $F_\ell$ appearing there:
\begin{eqnarray*}
&& F_5=k_5, \qquad  F_4 =k_4, \qquad F_3 =k_3+\frac52 k_5 V_5 ,\\
&& F_2 =k_2+2 k_4  V_5  -\frac{15}{4} i \hbar k_5  V_5' , \\
&& F_1 =k_1+\frac32 k_3 V_5   -2 i \hbar k_4 V_5'  + k_5 \left( \frac{15}{8}  V_5^2  -\frac{25}{8} \hbar^2 V_5''  \right), \\
&& F_0 =k_0+k_2 V_5-\frac{6}{8} i \hbar k_3 V_5'  +k_4 \left(V^2_5  - \hbar^2 V_5''  \right) - \frac{15}{16} i \hbar k_5 \left( 2 V_5  V'_5  +(i\hbar)^2 V'''_5 \right) ,
\end{eqnarray*}
where the $k_\ell$ are integration constants. The potential $V_5(x)$
must satisfy the following fifth order nonlinear differential
equation \be \label{quantpotn=5} {k_5} \left( \hbar ^4 V_5^{(v)} +
30 V_5^2  V_5' -20 \hbar ^2 V_5'  V_5''  -10 \hbar ^2 V_5  V_5'''
\right)+ {k_3} \left(24 V_5  V_5'   -4  \hbar ^2 V_5'''  \right) +16
{k_1} V_5' =16. \ee Some remarks are in order here: (i) if we take
$\hbar=0$ in this nonlinear fifth-order differential equation, we
recover the simple first order differential equation \eqref{bhcx5ev}
of the classical $n=5$ case; (ii) if we take $k_5=0$ we go back to
the case $n=3$ studied before \eqref{quantpotn=3}; (iii) only
constants $k_\ell$ with odd indices are present in
\eqref{quantpotn=5}: the constants $k_0, k_2$ and $k_4$ does not
play any role in the solution of the problem we are studying (in
particular, $k_0$ will be neglected in the sequel); (iv) the
equation can be integrated once to give \be \label{quantpotn=55}
k_5\left( \hbar ^4 V_5^{(iv)} + 10 V_5^3
 -5 \hbar ^2 (2  V_5 V_5''  +(V_5')^2 ) \right)+
{k_3} \left(
12 (V_5)^2  -4  \hbar ^2 V_5'' \right) +
16 {k_1} V_5 =16x+c_5.
\ee
As in the case $n=3$, the new relevant information of the present case is obtained by taking $k_1=k_3=0$ in \eqref{quantpotn=55}
\be\label{FV}
\hbar ^4 V_5^{(iv)}
 -10 \hbar ^2 \left(  V_5 V_5''  +\frac12 (V_5')^2 ) \right) + 10 V_5^3=\frac{16x+c_5}{k_5}.
\ee Remark that if $\hbar=0$, we recover basically the classical
$n=5$ result \eqref{n=5k_1=k_3=0vvv5}.  
With the simple transformations
\be
V_5= 2\hbar^2 \ W_5, \qquad x=z-\frac{c_5}{16},\qquad \kappa=\frac{8}{\hbar^6 k_5},
\ee
equation \eqref{FV} turns out to be
\be\label{CanonicalFV}
\frac{d^4 W_5}{dz^4}=20 W_5 \frac{d^2 W_5}{dz^2}+10\left( \frac{d W_5}{dz}\right)^2 -40 W_5^3+ \kappa z,
\ee
which appears  in the list of fourth order Painl\'eve equations of polynomial class, classified by Cosgrove \cite{Cosgrove2000}: it is precisely the so-called equation F-V (see Eq.~(2.67) of \cite{Cosgrove2000} with $\alpha=\beta=0$). Equation F-V has the Painl\'eve property and arises as group-invariant reduction of the KdV5 equation (a member of the KdV hierarchy). It is also a member of the so-called Painl\'eve-I hierarchy and is denoted by the notation $_{1}P_{4}$ \cite{Kudryashov1997}.  It is conjectured that 
F--V (in the nonautonomous case) defines a new transcendent in the sense that the general solution of  F--V cannot be expressed in terms of known transcendents including the six Painl\'eve transcendents, elliptic, hyperelliptic, abelian and automorphic functions.

The second order Painl\'eve transcendents $P_I$, $P_{II}$, $P_{IV}$ as quantum potentials have appeared previously \cite{Gravel2004,MarquetteWinternitz2008a,Hietarint1984, MarquetteWinternitz2007, Marquette2009}. A fourth order form of the potential equation which can be integrated in terms of solutions of the fourth Painl\'eve equation $P_{IV}$ first appeared in \cite{Gravel2004}.   To the best of our knowledge, the occurrence of a genuine fourth order Painl\'eve transcendent as potential is new. The surprising connection of superintegrability in the quantum case with soliton theory of infinite-dimensional integrable nonlinear systems manifests itself here once again.

The Heisenberg operator $\mathcal{B}_5$ has the form:
\begin{equation}\label{quantbn=5final}
\mathcal{B}_5=
\mathcal{B}_1+ k_2 \mathcal{H}_5 + k_3 \mathcal{B}_3^0  +k_4 \mathcal{H}_5^2
+ k_5 \mathcal{B}_5^0,
\end{equation}
where, the essentially new term is given by
\begin{equation}\label{quantbn=5finalzero}
\begin{array}{ll}
\mathcal{B}_5^0
&=P^5 + \frac52  V_5\,P^3  -\frac{15}{4} i \hbar  V_5' \,P^2 + \frac{5}{8}
\left( 3 V_5^2 -5 \hbar ^2 V_5'' \right)P - \frac{15}{16} i \hbar \left( 2 V_5 V'_5 -\hbar^2 V'''_5\right)
\\[2.ex]
&=\frac38(P^5 -\frac53(PHP^2 +P^2HP)+ \frac52(H^2P +P H^2))\,.
\end{array}
\end{equation}
Notice that the last expression is a symmetrized version of the corresponding
classical function (\ref{b50c}).


\item Case $n=6$

As we have already seen in the previous analysis, the even cases do not provide new information, and therefore we will skip the case $n=6$.


\item $n=7$

This is the last case that we will consider in this paper.
The Heisenberg operator is of
seventh order,
\begin{equation}\label{n7quantb7}
\mathcal{B}_7=
F_0 + F_1 \,P+ F_2 \,P^2+ F_3 \,P^3+ F_4 \,P^4+ F_5 \,P^5+ F_6 \,P^6+ F_7 \,P^7.
\end{equation}
Substituting (\ref{n7quantb7}) in (\ref{bhquant}) we get the following set of equations:
\begin{eqnarray}\label{efesquantn=7}
2 F_7'   \!\!&\!\!=\!\!&\!\!  0, \nonumber \\
2 F_6'  \!\!&\!\!=\!\!&\!\!  0 ,\nonumber  \\
2   F_5'   \!\!&\!\!=\!\!&\!\! 7  F_7 V_7',
\nonumber  \\
2 F_4'   \!\!&\!\!=\!\!&\!\! 6  F_6 V_7'+  i \hbar (F_5'' - 21 F_7 V_7'' ) ,  \nonumber  \\
2  F_3'   \!\!&\!\!=\!\!&\!\! 5 F_5 V_7' +  i \hbar (F_4''  -15 F_6 V_7'')  -35 \hbar ^2 F_7 V_7'''  ,
\nonumber  \\
2 F_2'   \!\!&\!\!=\!\!&\!\! 4  F_4 V_7'+   i \hbar( F_3''  -10 F_5 V_7'') -20 \hbar ^2 F_6 V_7''' + 35 i \hbar ^3 F_7 V_7^{(iv)}    ,
\\
2 F_1'   \!\!&\!\!=\!\!&\!\! 3 F_3 V_7' + i \hbar (F_2''  -6 F_4 V_7'') -10 \hbar ^2 F_5 V_7'''  +15 i \hbar ^3 F_6 V_7^{(iv)} +
21 \hbar ^4 F_7 V_7^{(v)} ,
\nonumber  \\
2 F_0'   \!\!&\!\!=\!\!&\!\! 2  F_2 V_7' + i \hbar (F_1'' -3 F_3 V_7'' ) - 4 \hbar ^2 F_4 V_7''' +5 i\hbar ^3 F_5 V_7^{(iv)} +6 \hbar ^4 F_6 V_7^{(v)}
-7i \hbar ^5 F_7 V_7^{(vi)} ,
\nonumber  \\
1  \   \!\!&\!\!=\!\!&\, F_1 V_7' +   i \hbar (F_0'' - F_2 V_7'' ) - \hbar^2  F_3 V_7'''  + i \hbar ^3 F_4 V_7^{(iv)} + \hbar ^4 F_5 V_7^{(v)} - i \hbar ^5 F_6 V_7^{(vi)}  -\hbar ^6 F_7 V_7^{(vii)} .  \nonumber
\end{eqnarray}
The solution of this system is:
\begin{eqnarray*}
F_7  \!\!&\!=\!&\!\! k_7, \ \ \  F_6 =k_6, \ \ \ F_5 =k_5+\frac72 k_7 V_7 , \ \ \ F_4 =k_4+3 k_6  V_7  -\frac{35}{4} i \hbar k_7  V_7' , \\
F_3  \!\!&\!=\!&\!\! k_3 +\frac52 k_5 V_7   -6 i \hbar k_6 V_7'  + \frac{35}{8}  k_7 \left( V_7^2  -3 \hbar ^2 V_7''  \right), \\
F_2  \!\!&\!=\!&\!\! k_2 +2 k_4 V_7
-\frac{15}{4} i \hbar {k_5}   V_7'  +k_6 \left(3 V^2_7  - 7\hbar^2 V_7''  \right) - \frac{35}{16} i \hbar k_7 \left( 6 V_7  V'_7  +5(i\hbar)^2 V'''_7 \right) ,
\\
F_1  \!\!&\!=\!&\!\! k_1 +\frac{3}{2} k_3 V_7 -2 i k_4 \hbar  V_7' +\frac{5}{8}k_5 \left(3 V_7^2-5 \hbar ^2 V_7'' \right)+ i \hbar k_6 \left(4  \hbar ^2 V_7''' -6  V_7  V_7' \right)+\\ \nonumber
& \!&\!\!  +\frac{1}{32} k_7 \left({161}\hbar ^4 V_7^{(iv)} - 350 \hbar ^2 V_7  V_7'' - {245} \hbar ^2 (V_7')^2+70 V_7^3\right),
\\
F_0  \!\!&\!=\!&\!\! k_0 +k_2 V_7 -\frac{3}{4} i k_3 \hbar  V_7' +k_4 \left(V_7^2-\hbar ^2 V_7'' \right)+\frac{15}{16} i \hbar k_5 \left(\hbar ^2 V_7''' - 2 V_7  V_7' \right)+ \\  \nonumber
& \!&\!\!  +k_6 \left(\hbar ^4 V_7^{(iv)} -3 \hbar ^2 V_7  V_7'' -2 \hbar ^2 (V_7')^2+V_7^3\right)\\  \nonumber
& \!&\!\!
+ \frac{21}{64}  i \hbar k_7
\left(-3 \hbar ^4 V_7^{(v)}  +10 \hbar^2 V_7  V_7''' - 10 V_7^2 V_7' +
20 \hbar ^2 V_7'  V_7'' \right),
\end{eqnarray*}
where the $k_\ell$ are integration constants, and the potential $V_7 $ must satisfy the following seventh order nonlinear differential equation:
\begin{eqnarray}\label{quantpotn=7}
k_7 \Bigl[
\hbar ^6 V_7^{(vii)} -14 \hbar^4 (V_7  V_7^{(v)}  +3 V_7^{(iv)}  V_7' +5 V_7''' V_7'' )+70 \hbar ^2 (V_7 ^2 V_7''' + (V_7')^3 +4 V_7  V_7'  V_7'')
 \nonumber
\\
-140 V_7^3 V_7'\Bigr]
 +4 {k_5} \left[- \hbar ^4 V_7^{(v)}  + 10 \hbar ^2 (V_7  V_7''' + 2 V_7'  V_7'' )-30 V_7 ^2 V_7'  \right]
 \nonumber
  \\
+ 16 {k_3} \left[ \hbar ^2 V_7'''  -6 V_7  V_7'  \right]
-64 k_1 V_7' +64=0.
\end{eqnarray}
Remark that only $k_\ell$ with odd indices are present: the constants $k_0, k_2, k_4$ and $k_6$ do not play any role in the solution of the problem we are studying (indeed, as in the previous cases, we will omit $k_0$ from now on). If we consider the limit case $\hbar=0$, the nonlinear differential equation \eqref{quantpotn=7} reduces to the much simpler equation \eqref{bhcx77ev}. In spite of its formidable aspect, equation \eqref{quantpotn=7} can be integrated once:
\begin{eqnarray}\label{quantpotn=7integrated}
k_7 \left[ -\hbar ^6 V_7^{(vi)}
+7 \hbar^4 (2 V_7  V_7^{(iv)}  +4 V_7''' V_7' +3 (V_7'')^2 )
-70 \hbar^2 (V_7^2 V_7'' + (V_7')^2 V_7) +35 V_7^4\right]
\!\!\!\!\!\!&\!\!\!\!\!\!\!&\!\!\!\!\!\!
  \\
+4k_5\left[\hbar ^4 V_7^{(iv)} -5 \hbar ^2 (2V_7 V_7'' +(V_7')^2)+10 V_7^3 \right] -16k_3\left[\hbar ^2 V_7'' -3(V_7)^2\right]+64 k_1 V_7 \!\!\!&\!\!= \!\!&\!\! 64x+c_7. \nonumber
\end{eqnarray}
If we consider here $\hbar=0$, all the terms with derivatives of
$V_7(x)$ completely disappear, and the corresponding fourth order
polynomial equation \eqref{bhcx77evb} is obtained for the classical
potential.

The special case of \eqref{quantpotn=7integrated} where $k_1=k_3=k_5=0$ and $k_7\ne 0$ gives rise to a novel potential
\begin{equation}\label{quantpotn=7painleve}
\hbar ^6 V_7^{(vi)} -7 \hbar^4 \Bigl(2 V_7  V_7^{(iv)}  +4 V_7'''
V_7' +3 (V_7'')^2 \Bigr) +70 \hbar^2  \Bigl(V_7^2 V_7'' + (V_7')^2
V_7  \Bigr) -35 V_7^4=\frac{64x+c_7}{-k_7}.
\end{equation}
We have checked if  Eq. \eqref{quantpotn=7painleve}  passes the
Painlev\'{e} test, what is only a necessary condition for the
equation to possess the Painlev\'{e} property. The resonances occur
at $r=2,4,5,7,10$ at which all compatibility conditions are
satisfied which implies that the test is passed. This feature is
typical of all quantum potentials obtained so far.  
The simple transformations 
\be
V_7(x)=\hbar^2 W_7(x), \qquad x=z-\frac{c_7}{64}, \qquad \kappa=-\frac{64}{k_7 \hbar^8},
\ee
allows to transfer the dependence on $\hbar$ of the whole equation to the independent variable, transforming Eq. \eqref{quantpotn=7painleve} into
\be
W_7^{(vi)}= 14 W_7  W_7^{(iv)}  +28 W_7''' W_7' +21 (W_7'')^2 
-70  W_7^2 V_7'' -70 (W_7')^2 W_7 +35 W_7^4+\kappa z.
\ee
As we have already mentioned, this sixth-order nonlinear differential equation correspond to some sixth-order Painlev\'{e} equation to be determined.

The integral of motion $\mathcal{B}_7$ has the form:
\begin{equation} 
\mathcal{B}_7 =
k_1 P + k_2 \mathcal{H}_7 + k_3\mathcal{B}_3^0  + k_4\mathcal{H}_7^2 + k_5\mathcal{B}_5^0+ k_6 \mathcal{H}_7^3 + k_7 \mathcal{B}_7^0,
\end{equation}
where, again, new information comes only from the last term:
\begin{eqnarray*}
\mathcal{B}_7^0 \!\!&\!\!=\!\!&\!\! P^7  +\frac72  V_7P^5 -\frac{35}{4} i \hbar  V_7' P^4  + \frac{35}{8}  \left( V_7^2  -3 \hbar ^2 V_7''  \right)  P^3  - \frac{35}{16} i \hbar \left( 6 V_7  V'_7  -5\hbar^2 V'''_7 \right)\,P^2  \\
 &&+
\frac{1}{32} \left({161}\hbar ^4 V_7^{(iv)} - 350 \hbar ^2 V_7  V_7'' - {245} \hbar ^2 (V_7')^2+70 V_7^3\right)\,P \\
 && - \frac{21}{64}  i \hbar  \left(3 \hbar ^4 V_7^{(v)}  -10 \hbar ^2 V_7  V_7''' + 10 V_7^2 V_7' - 20 \hbar ^2 V_7'  V_7'' \right).
\end{eqnarray*}

\end{itemize}

\subsection{Superintegrable Hamiltonians and their symmetries}

Once we have the Heisenberg Hamiltonians $\langle \mathcal{H}_{mx}, \mathcal{B}_{mx}, 1\rangle$,
and $\langle \mathcal{H}_{ny}, \mathcal{B}_{ny}, 1\rangle$ we can write a superintegrable
Hamiltonian by adding them just as in the classical context:
\begin{equation}
\mathcal{H}_{mn} = \mathcal{H}_{mx} + \mathcal{H}_{ny},
\qquad V_{mn}(x,y) = V_{mx}(x) + V_{ny}(y),
\end{equation}
where the degrees $m$ and $n$ of each Hamiltonian can be different.
The third symmetry of odd degree given by ${\rm max}(m,n)$ is
\begin{equation}
\mathcal{B}_{mn} = \mathcal{B}_{mx} - \mathcal{B}_{ny}\,.
\end{equation}
In this way we can trivially extend this method to get superintegrable
Hamiltonians in three or higher dimensions.

However, the one dimensional potentials involved are given by solutions of non linear
differential equations that are not well known in much detail.
For instance, some of the potentials may have singularities or may
not be well defined in the whole real line. This can depend on
the initial conditions imposed to the solutions.
In this sense the study of all the possible potentials is as complex
as the classification of the solutions of such non--linear equations.

In the quantum case the pasting of `local' superintegrable
potentials in order to get a potential defined in the whole plane is
out of place. For instance, in the same way as (\ref{vt11}), we
could define the potential
\begin{equation}\label{vt11q}
\tilde V_{11} = |x| + |y|\,.
\end{equation}
But we can not apply any property related to superintegrability
to this two--dimensional system. Here, we can not act as in the classical case,
pasting the trajectories of different domains.

\section{Conclusions}

In this work we have carried out a systematic study of superintegrable
Hamiltonian systems separable in Cartesian coordinates such that each
component is of Heisenberg type. A one-dimensional Hamiltonian
$H_n$ is said to be of Heisenberg type in the classical context
if there is a function
$B_n$ of degree $n$ in the momentum variable such that the Heisenberg Poisson commutator
$\{H_n, B_n\} = 1$ is satisfied. In the quantum frame a similar definition applies
for a Hamilton operator ${\cal H}_n$ and a polynomial operator ${\cal B}_n$
that satisfy
the commutator $[{\cal H}_n,{\cal B}_n] = i\hbar$.

In the classical case we have found a general solution to this
problem for any value of $n$. The relevant solutions are realized
for the odd values $n = 2\ell +1$. The potentials of this type of
Hamiltonians satisfy an algebraic equation of degree $\ell+1$. A
representative potential for such a value is given by a root of
index $\ell+1$: $ v_{2\ell+1}(x)\propto  x^{1/(\ell+1)}$, $\ell\in
\mathbb N$. Some of the resulting superintegrable Hamiltonians are
defined in a region of the plane (in the case of two cartesian
coordinates) so that we are lead to a restricted concept of
superintegrability. This type of potentials  do not allow for
classical bounded motions, so that a particle that initially is in
one of these regions, in general after a time  will leave it and
cross to another region where the superintegrability is not
satisfied. In conclusion, for some cases we can describe only a part
of the motion by means of the superintegrability properties for such
a kind of Heisenberg systems.

In the quantum case it is possible to work out the solutions for any
value of $n$, but we have not found closed expressions. It is shown
that, as in the classical case, the odd values $n = 2\ell +1$ are
relevant. For some values of $\ell$, the expression for the operator
${\cal B}_{2\ell+1}$ in terms of the potential function
$V_{2\ell+1}(x)$ has been explicitly computed, as well as the
differential equation that  $V_{2\ell+1}(x)$ must satisfy. Contrary
to the corresponding classical analog, here the  equations (except
for the case $\ell=0$ that can be integrated) are not algebraic, but
nonlinear differential equations that can not be integrated in terms
of known special functions. In fact, they belong to a type of higher
order Painlev\'e equations, starting with Painlev\'e I for $\ell =
1$. Given the equations for the potential and the expressions for
the Heisenberg operators of the quantum problem in terms of the
potential, then if we perform the classical limit $\hbar \to 0$, the
corresponding classical equations for $v_{2\ell+1}(x)$, as well as
the classical expressions $B_{2\ell+1}$  are recovered, in a certain
sense.

Some particular solutions of the general approach contained in this paper can
be found in previous references \cite{Gravel2004, MarquetteWinternitz2007, MarquetteWinternitz2008a}. 
For example, in  \cite{Gravel2004} 
the symmetries  of the two-dimensional
Euclidean systems separable in Cartesian coordinates, up to third order, 
are exhaustively studied;
the results  include as particular cases all our solutions up to
order three: In the quantum systems these potentials are labeled as (Q.17) and
(Q.20), while in the classical framework are the cases (C.5) and (C.7).
Reference \cite{MarquetteWinternitz2007} analyses the same problem as \cite{Gravel2004}  paying attention to the algebraic structure of the symmetries. 
Some of the potentials they obtained (cases 5, 7 and 8, where the symmetry algebra is Heisenberg) are the same as in our work.
We have carefully explained that the local superintegrability affect the trajectories corresponding to these three cases. Another reference dealing with a similar strategy is \cite{Marquette2011}, where the author is also searching for
higher order symmetries, for the same type of systems, by means of ladder operators. However, 
the difference is that we use as the basic ingredient 
the Heisenberg algebra instead of the ladder algebra.

In conclusion, we have shown here that a fruitful way to find higher order
symmetries of classical and quantum systems can be based on the algebraic
properties of the corresponding Hamiltonian. In the present work this key
idea is successfully implemented by looking for superintegrable Hamiltonians
of Heisenberg type.

\bigskip

\subsection*{ {Acknowledgments}}

We acknowledge the financial support of the Spanish MINECO (Project MTM2014-57129-C2-1-P) and Junta de Castilla y
Le\'on (VA057U16). F.~G\"{u}ng\"{o}r and \c{S}.~Kuru acknowledge the warm hospitality
during their stays at Department of Theoretical Physics, University
of Valladolid, Spain.

\end{document}